\documentclass{webofc}

\usepackage[varg]{txfonts}   
\usepackage{hyperref}
\usepackage{url}
\usepackage{amsmath, amssymb, bm, physics, mathrsfs}
\usepackage{booktabs}

\usepackage{graphicx}
\usepackage{xcolor}
\hypersetup{colorlinks=true,citecolor=blue,urlcolor=blue,linkcolor=blue}

\newcommand{\mA}{$m_{A'}$}

\begin{document}
\title{Dark photons in exotic Higgs boson decays at FCC-ee}

\author{\firstname{Axel} \lastname{Gall\'en}\inst{1}\fnsep\thanks{\email{axel.lars.gallen@cern.ch}} \and
        \firstname{Sarah} \lastname{Ben Abdesselem}\inst{1} \and
        \firstname{Giulia} \lastname{Ripellino}\inst{1} \and 
        \firstname{Rebeca} \lastname{Gonzalez Suarez}\inst{1}
}

\institute{Department of Physics and Astronomy, Uppsala University,
Lägerhyggsvägen 1, Uppsala, 752 37, Sweden
}

\abstract{
A prospective search for dark photons decaying into collimated, displaced muon pairs is presented for the electron–positron stage of the Future Circular Collider, FCC-ee. Assuming the full luminosity of the Higgs run at 240~GeV, the study probes a broad range of dark photon masses and couplings. The signature comprises two jets from a Z boson decay and two muon pairs from dark photons produced in Higgs decays. A kinematics-based selection achieves zero background while retaining high signal efficiency. Expected limits are set on the Higgs boson branching ratio to dark photons as a function of the dark photon mass and coupling. 
}

\maketitle

\section{Introduction}

\label{intro}
The Standard Model (SM) of particle physics is not yet able to answer several open questions, such as the nature of dark matter (DM), the origin of neutrino masses, or the baryon asymmetry of the Universe. A broad class of beyond the SM (BSM) theories addressing these issues proposes hidden sectors, consisting of new particles and forces, parallel to the SM but that interact almost imperceptibly with it via portal interactions. The spin of the mediator particle determines the type of portal considered (fermion, scalar, pseudoscalar). Hidden sectors related to DM are commonly referred to as dark sectors. In this work, we focus on the vector portal and propose a prospective search for dark photons.

Among the various dark photon scenarios, we focus on the Hidden Abelian Higgs Model (HAHM)~\cite{Curtin:2013fra}, in which the SM is extended by a hidden $U(1)_D$ gauge symmetry and an additional Higgs field. Upon spontaneous symmetry breaking, the corresponding gauge boson acquires a mass, giving rise to a new vector state, the dark photon $A'$. The model is parametrized by the kinetic mixing $\varepsilon$, the dark photon mass \mA, the Higgs portal coupling $\kappa$, and the dark Higgs mass $m_s$. The kinetic mixing parameter and the dark photon mass determine the lifetime of the dark photon, which scales as $\varepsilon^{-2}m_{A'}^{-1}$. For sufficiently small $\varepsilon$ and \mA, the dark photon can acquire a macroscopic decay length, leading to displaced signatures. The Higgs portal coupling $\kappa$ governs the mixing between the SM Higgs and the dark Higgs inducing a coupling between the SM Higgs boson $h$ and pairs of dark photons. If kinematically allowed, this opens the exotic decay channel $h \to A'A'$, providing a direct probe of hidden-sectors at Higgs factories.

The Future Circular Collider (FCC) is a proposed next-generation collider at CERN~\cite{FCC:2025lpp}, designed to succeed the Large Hadron Collider (LHC)~\cite{European:2957411}. With a circumference of approximately 91~km, the FCC will collide $e^+ e^-$ (FCC-ee) and later hadrons (FCC-hh). FCC-ee will provide a clean experimental environment with very high luminosity, enabling precision measurements and sensitive searches for rare processes. In particular, the $Zh$ run at $\sqrt{s}=240~\text{GeV}$ is expected to produce more than 2 million Higgs bosons. This makes FCC-ee an ideal facility for both precision Higgs studies and searches for exotic Higgs decays.

\begin{figure}[ht!]
\centering
 \includegraphics[width=0.4\linewidth]{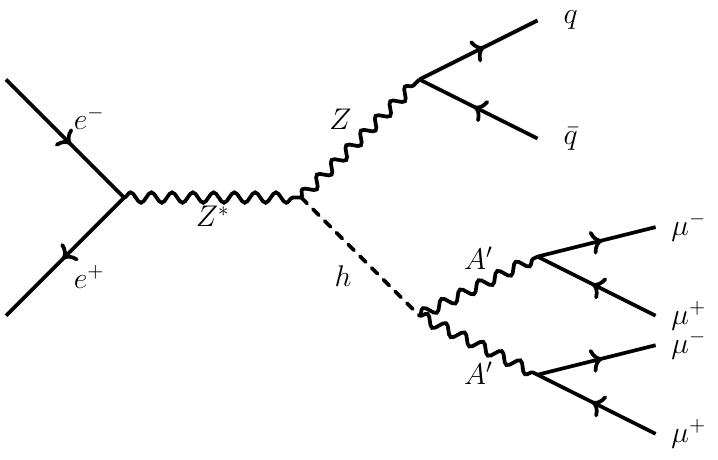}
 \caption{Feynman diagram for the considered signal process.}
\label{fig:eeZh}
\end{figure}

In this paper, we present a sensitivity study of the exotic decay $h \to A'A'$, with each dark photon decaying into a pair of muons, as illustrated in Fig.~\ref{fig:eeZh}. The process is studied at the $Zh$ run at $\sqrt{s}=240~\text{GeV}$, assuming an integrated luminosity of $L=10.8~\text{ab}^{-1}$.

This analysis targets hadronic $Z$ decays, producing a signature characterized by two jets, and two pairs of muons, each pair originating from a dark photon decay. The scenario corresponds to a search for long-lived particles (LLPs)~\cite{Alimena:2019zri}, as the small couplings considered lead to displaced, collimated muon pairs.

Previous studies of dark photons at FCC-ee can be found in~\cite{Polesello:2025cbi} and are included in the global assessment of~\cite{deBlas:2025gyz}.

\section{Signal and Background Samples}
Signal samples are generated using the \texttt{HAHM\_MG5model\_v3 UFO} model~\cite{Curtin:2014cca} within \textsc{MadGraph5\_aMC@NLO v3.5.11}~\cite{Alwall:2011uj}. Parton showering and hadronization are performed with \textsc{Pythia8}~\cite{Sjostrand:2014zea}, and the detector response is simulated using the \textsc{Delphes}~\cite{deFavereau:2013fsa} fast simulation with the IDEA detector card~\cite{Ilg:2025zri}, corresponding to the FCC-ee \texttt{Winter2023} campaign~\cite{WINTER2023}.

The samples are generated at $\sqrt{s}=240$~GeV, with 10k events per signal point. The signal grid spans $(m_{A'},\varepsilon)$ values with $m_{A'}$ in the range $0.36 - 10$~GeV and $\varepsilon$ between $8.7\times10^{-8}$ and $6.7\times10^{-5}$, corresponding to approximately 1300 parameter points. This setup probes mean proper lifetimes ranging from $\mathcal{O}(\mu\text{m})$ to $\mathcal{O}(\text{m})$. The grid is optimized based on a preliminary study of the parameter space in which dark photons decay within the tracker volume of the IDEA detector.

In this study, the dark scalar is assumed to be heavy ($m_s=200$~GeV) and weakly coupled to the SM Higgs ($\kappa=10^{-3}$ ), suppressing its direct phenomenological impact. Under these assumptions, the branching ratio for Higgs decays to dark photons depends primarily on the dark photon mass, while remaining sizable across the mass range considered in this analysis. A detailed description of the model and its phenomenology can be found in~\cite{Curtin:2014cca}.

All signal event yields are normalized via the following in order to match the number of expected events at the FCC-ee $Zh$ pole:
\begin{equation}\label{eq:Nevents}
    N_{events} = N_{Zh} \times \mathcal{B}(Z \to q\overline{q}) \times \mathcal{B}(h\to A'A') \times  \mathcal{B}(A' \to \mu^+ \mu^-)^2
\end{equation}
where $N_{Zh}=2.6\times10^6$ is the amount of expected events at the $Zh$ pole~\cite{Dam:2025zed}, $\mathcal{B}(Z \to q\overline{q})$ is obtained from~\cite{ParticleDataGroup:2024cfk} and $\mathcal{B}(A' \to \mu^+ \mu^-)$ is obtained from~\cite{Curtin:2014cca}. The branching ratio for Higgs decays to dark photons $\mathcal{B}(h\to A'A')$ is varied in the interpretation of the results.

The dominant SM background processes considered are those that can either mimic the signal signature or have sufficiently large cross sections to contribute non-negligibly. The only SM process yielding the same final state is Higgs boson decays to two Z bosons which subsequently decay into two muons each. In addition, inclusive $ZZ$ and $WW$ production is considered due to their large cross sections. While the signal samples are produced privately, the background samples are taken from the official \texttt{winter2023} central production campaign~\cite{WINTER2023}. The corresponding samples are summarized in Table~\ref{tab:backgrounds}.

\begin{table}[ht!]
  \centering
  \caption{Background samples used in the analysis. All the samples come from the \texttt{winter2023} production campaign.}
  \label{tab:backgrounds}
  \begin{tabular}{llr}
    \toprule
    Process & Sample name & $N_\mathrm{events}$ \\
    \midrule
    $e^+e^-\to Zh$, $Z\to q\bar{q}$, $h\to ZZ\to 4\ell$ & \texttt{wzp6\_ee\_qqH\_HZZ\_llll\_ecm240} & 1\,200\,000 \\
    $e^+e^-\to ZZ$ & \texttt{p8\_ee\_ZZ\_ecm240} & 49\,773\,829 \\
    $e^+e^-\to WW$ & \texttt{p8\_ee\_WW\_ecm240} & 374\,030\,096 \\
    \bottomrule
  \end{tabular}
\end{table}

\section{Object Definition}
This analysis is performed using the \textsc{FCCAnalyses} framework~\cite{FCCANALYSES}, in the \texttt{pre-edm4hep1} branch compiled against the 2024-03-10 Key4hep stack~\cite{Carceller:2025fjc}, ensuring compatibility with the FCC-ee \texttt{winter2023} centrally produced samples. The Delphes output is converted to the EDM4HEP data format using the \texttt{k4SimDelphes} package and processed with ROOT RDataFrame as the event processing engine.

The analysis considers jets and muons. Jets are clustered using the exclusive Durham-$k_t$ algorithm~\cite{Catani:1991hj}, as implemented in the \textsc{FastJet} interface of \textsc{FCCAnalyses}. The clustering proceeds by iteratively merging particle pairs according to the distance measure
\begin{equation}
    d_{ij} = 2\min\qty(E_i^2, E_j^2)\qty(1 - \cos\theta_{ij}),
\end{equation}
where $E_i$ and $E_j$ are the particle energies and $\theta_{ij}$ is their opening angle. The procedure is stopped when exactly two jets are reconstructed, targeting the hadronic decay $Z \to q\bar{q}$ and implying a requirement for the selected events to have at least two jets. A minimum transverse momentum requirement of $p_T \geq 5~\text{GeV}$ is imposed on both jets.

Muon candidates are required to satisfy $p_T > 0.1$~GeV and to lie within the acceptance of the tracking volume. The muon momentum is taken directly from the associated muon track. Signal muons originate from displaced decays. To reconstruct displaced vertices (DVs), the \texttt{VertexFitter\_Tk} algorithm from the \texttt{VertexFitterSimple} class in LCFIPlus is used~\cite{LCFIplus}. The algorithm fits a common secondary vertex to the muon tracks, providing the three-dimensional vertex position and its covariance matrix. The decay length $L_{xyz}$ of each dark photon candidate is then computed as the distance between the reconstructed secondary vertex and the primary interaction vertex.

\section{Event Selection}
A preselection is defined to restrict the signal phase space. Events are required to contain at least four reconstructed muons, forming a $\mu^+\mu^-\mu^+\mu^-$ final state, and at least two jets with $p_T \geq 5~\text{GeV}$. 

This preselection removes more than 99\% of the total background while retaining approximately 90\% of the signal across the parameter space. A final kinematic selection is then applied to further refine the search and reject the remaining background.

The muons are then grouped into two opposite-sign pairs by minimizing the angular separation $\Delta R$ between muons of opposite charge, selecting the pairing that yields the smallest $\Delta R$ values.

Since the four muons in the signal events originate initially from a Higgs boson decay, the invariant mass of the four muons is required to lie within the Higgs boson mass window, $M_{4\mu} \in [120,130]~$GeV.

Each muon pair is assumed to originate from a dark photon decay. The invariant mass of each pair is therefore required to be $M_{\mu\mu} < 11~$GeV. This reflects the mass range considered for the dark photon signal and suppresses contributions from higher-mass resonances.

To select displaced decays, a secondary vertex is reconstructed from the two tracks corresponding to each di-muon pair. Events are then required to contain exactly two DVs with the distance between the primary and secondary vertex in the range $L_{xyz} \in [3,2000]~$mm. This targets the long-lived regime and rejects the remaining prompt SM background.

The full preselection and kinematic selection are presented in Tab.~\ref{tab:selection_summary}.

\begin{table}[ht!]
\centering
\caption{Summary of the criteria defining the preselection and kinematic selection.}
\begin{tabular}{ll}
\toprule
\multicolumn{2}{l}{\textit{Preselection}} \\
\midrule
Number of jets ($p_T \geq 5$~GeV) & $\geq 2$  \\
Number of muons ($p_T > 0.1$~GeV) & $\geq 4$ \\
Charge requirement & $N_{\mu^+} \geq 2$, $N_{\mu^-} \geq 2$ \\
\midrule
\multicolumn{2}{l}{\textit{Kinematic selection}} \\
\midrule
Muon pairing based on min. $\Delta R$ &  \\
$M_{4\mu}$ & $\in [120, 130]$ GeV \\
$M_{\mu\mu}$ & $< 11$ GeV \\
$L_{xyz}$ & $\in [3, 2000]$ mm \\
\bottomrule
\end{tabular}
\label{tab:selection_summary}
\end{table}

Three benchmark samples are selected for illustration purposes. The first sample is generated with $m_{A'} = 0.36$~GeV and $\varepsilon = 1.00 \times 10^{-5}$; the second sample with $m_{A'} = 1.1$~GeV and $\varepsilon = 3.12 \times 10^{-6}$; and the third with $m_{A'} = 7.0$~GeV and $\varepsilon = 2.70 \times 10^{-7}$. These samples offer a good range of dark photon masses and couplings, but also lifetime and proper decay length, covering distances from about 2~mm (first sample) to 148~mm (third).

\begin{figure}[ht!]
\centering
 \includegraphics[width=0.49\linewidth]{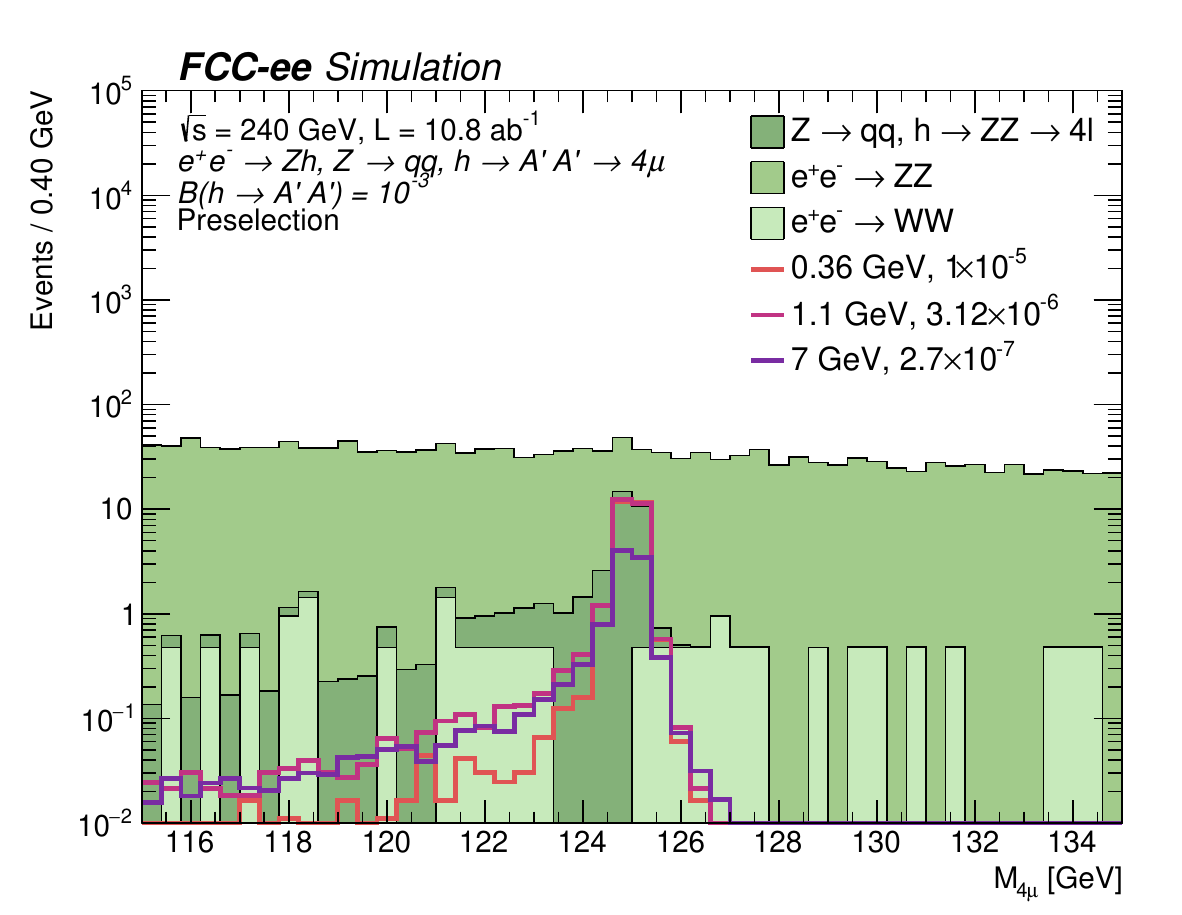}
  \includegraphics[width=0.49\linewidth]{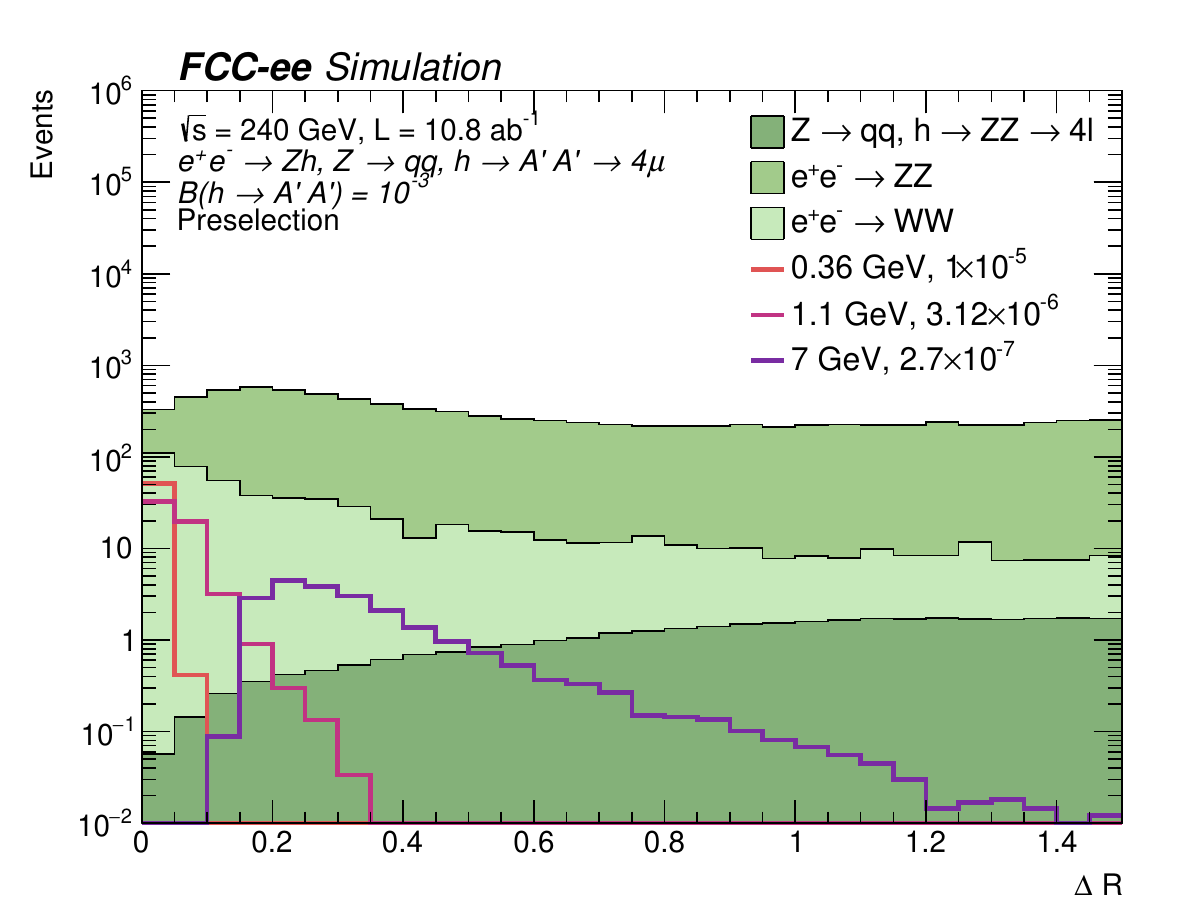}
  \includegraphics[width=0.49\linewidth]{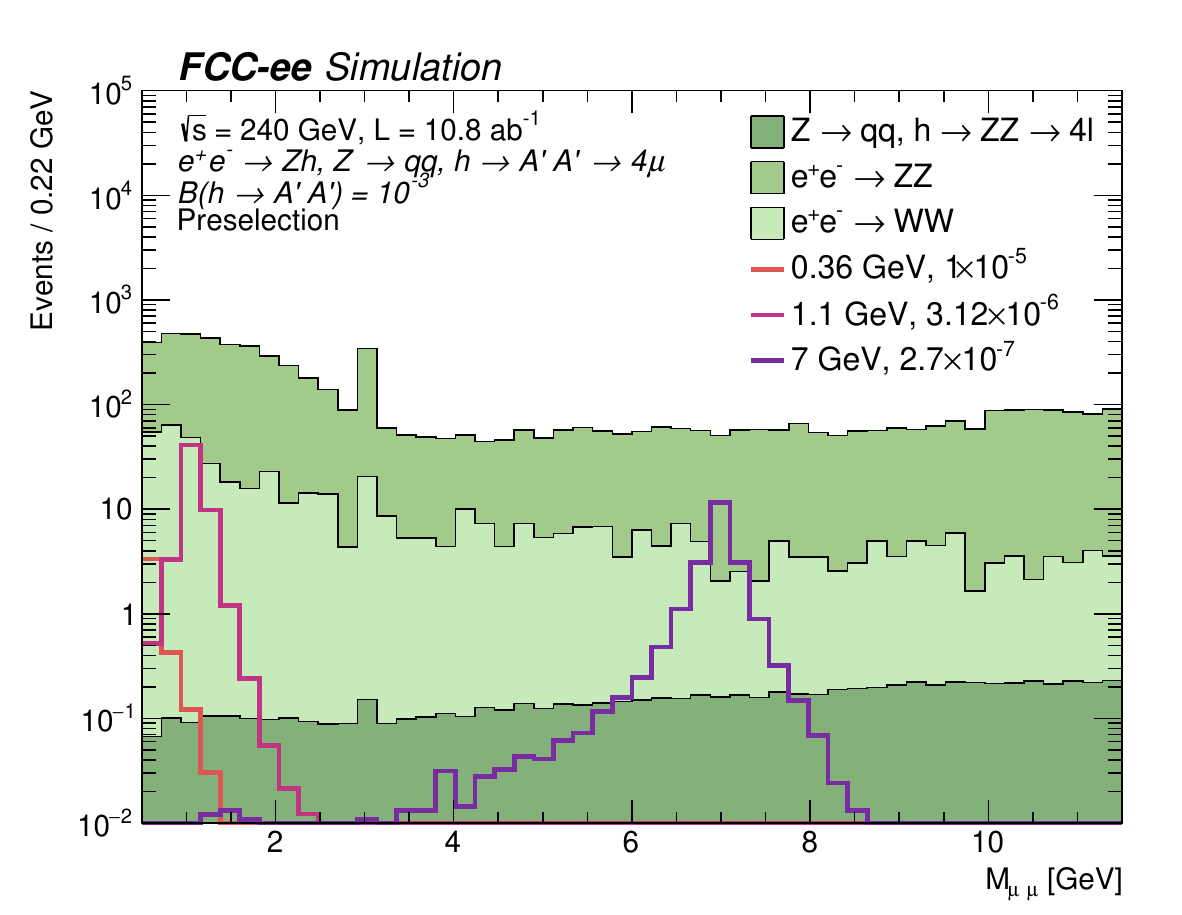}
  \includegraphics[width=0.49\linewidth]{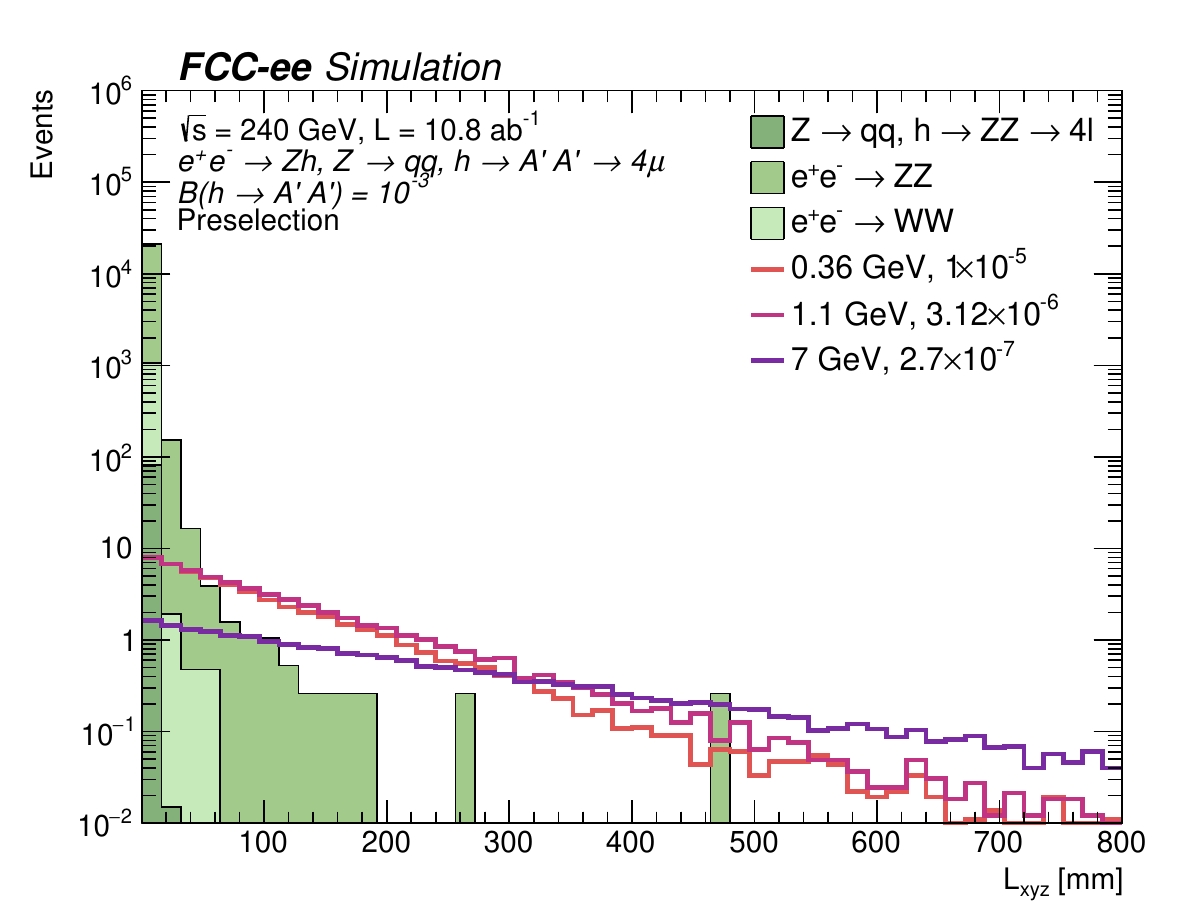}
 \caption{Kinematic distributions of the variables used in the final selection for three benchmark signals and backgrounds after preselection: (Top Right) Invariant mass of the four-muon system; (Top Left) $\Delta R$ of the muon pairs; (Bottom Right) Invariant mass of the muon pairs; (Bottom Left) $L_{xyz}$ of the di-muon vertices. Background samples are normalized to the expected luminosity, and the signal is normalized according to Eq.~\ref{eq:Nevents} assuming $\mathcal{B}(h\to A'A')=10^{-3}$.}
\label{fig:kinematics}
\end{figure}

Kinematic distributions of the variables used in the final selection for the three benchmark signals and backgrounds are presented after preselection in Fig.~\ref{fig:kinematics}.

These selections reduce the SM background to zero while retaining a significant fraction of signal events. The number of events after the full selection is presented in Table~\ref{tab:cutflowbck} for the backgrounds, and in Table~\ref{tab:cutflow} for the three representative signal points and the sum of all backgrounds.

\begin{table}[ht!]
\centering
\caption{Expected event yields after each selection step for the backgrounds, normalized to the expected luminosity. Uncertainties are only statistical.}
  \label{tab:cutflowbck}
    \resizebox{\textwidth}{!}{%
\begin{tabular}{l|c|c|c|c}
\toprule
 & $\mathbf{h \rightarrow ZZ \rightarrow 4\ell}$ & \textbf{ZZ} & \textbf{WW} & \textbf{Sum} \\
\hline
Before selection & $407.92 \pm 0.37$ & $14\,677\,100 \pm 2\,000$ & $177\,535\,800 \pm 9\,200$ & $192\,213\,300 \pm 9\,400$ \\
Preselection & $45.86 \pm 0.12$ & $11\,000 \pm 54$ & $518 \pm 16$ & $11\,561 \pm 56$ \\
\hline
$120 < M_{4\mu} < 130$ GeV & $34.45 \pm 0.11$ & $815 \pm 15$ & $8.6 \pm 2.0$ & $858 \pm 15$ \\
$M_{\mu\mu} < 11$ GeV & $0.09 \pm 0.01$ & $2.35 \pm 0.78$ & $0.48 \pm 0.48$ & $2.91 \pm 0.92$ \\
$3\,\mathrm{mm} < L_{xyz} < 2000\,\mathrm{mm}$ & $\leq 0.02$ & $\leq 1.6$ & $\leq 0.96$ & $\leq 1.8$ \\
\hline
Total Selection & $0$ & $0$ & $0$ & $0$ \\
\bottomrule
  \end{tabular}}
\end{table}

\begin{table}[ht!]
\centering
\caption{Expected event yields after each selection step for the three benchmark signal samples, denoted by $(m_{A'},\varepsilon)$, normalized according to Eq.~\ref{eq:Nevents} assuming $\mathcal{B}(h\to A'A')=10^{-3}$, presented together with the total SM background, normalized to the expected luminosity. Uncertainties are only statistical.}
  \label{tab:cutflow}
    \resizebox{\textwidth}{!}{%
\begin{tabular}{l|c|c|c|c}
\toprule
 & \multicolumn{3}{c|}{\textbf{Signal}} & \textbf{Background} \\
& 0.36~GeV, $1\times 10^{-5}$ & 1.1~GeV, $3.12\times 10^{-6}$ & 7~GeV, $2.7\times 10^{-7}$ & \\
\hline
Before selection & $285.4 \pm 2.9$ & $159.4 \pm 1.6$ & $29.98 \pm 0.30$ & $192\,213\,300 \pm 9400$ \\
Preselection & $266.6 \pm 2.8$ & $148.0 \pm 1.5$ & $27.19 \pm 0.29$ & $11\,561 \pm 56$ \\
\hline
$120 < M_{4\mu} < 130$ GeV & $264.7 \pm 2.8$ & $142.9 \pm 1.5$ & $24.92 \pm 0.27$ & $858 \pm 15$ \\
$M_{\mu\mu} < 11$ GeV & $264.7 \pm 2.8$ & $142.9 \pm 1.5$ & $24.91 \pm 0.27$ & $2.91 \pm 0.92$ \\
$3\,\mathrm{mm} < L_{xyz} < 2000\,\mathrm{mm}$ & $248.3 \pm 2.7$ & $134.9 \pm 1.5$ & $24.24 \pm 0.27$ & $\leq 1.8$ \\
\hline
Total Selection & $248.3 \pm 2.7$ & $134.9 \pm 1.5$ & $24.24 \pm 0.27$ & $0$ \\
\bottomrule
  \end{tabular}}
\end{table}

\section{Results}
The full event selection successfully reduces the background to zero while retaining a significant number of events for representative signal points.

Applying the selection to the full signal grid allows us to set preliminary limits on the branching ratio $\mathcal{B}(h\to A'A')$ in the plane defined by the dark photon mass, $m_{A'}$, and the kinetic mixing parameter, $\varepsilon$.

In the zero-background scenario, the sensitivity is determined using Poisson statistics. For an observation compatible with zero events, the 95\% confidence level (CL) upper limit on the signal yield is approximately 3 events. This defines the exclusion threshold used in this analysis.

\begin{figure}[ht!]
    \centering
    \includegraphics[width=0.8\linewidth]
    {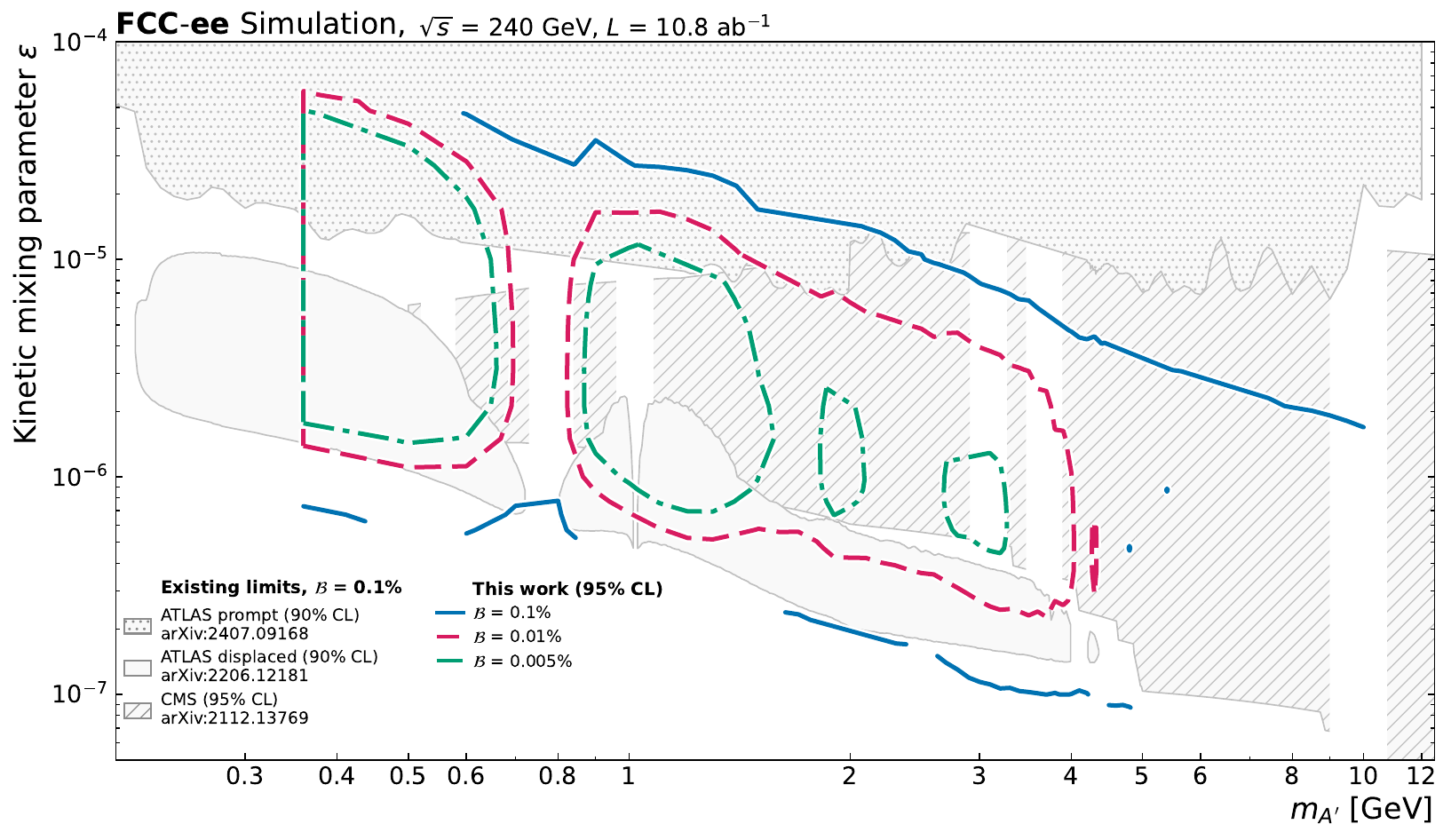}
    \includegraphics[width=0.8\linewidth]
    {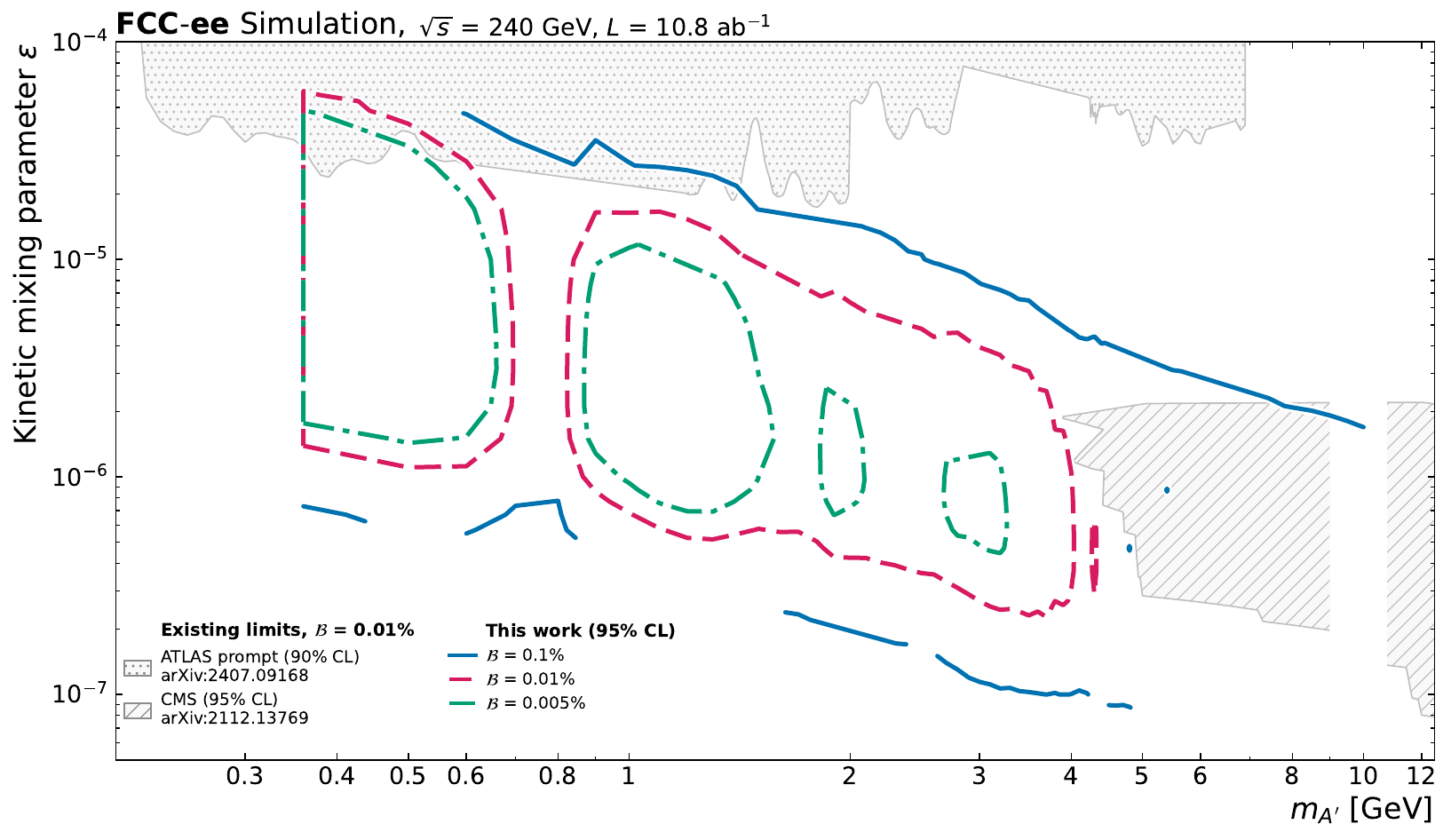}
    \caption{Exclusion limits at 95\% CL obtained in this analysis for $\mathcal{B}(h\to A'A') = 0.1\%$ (blue), $0.01\%$ (red), and $0.005\%$ (green). The results are compared to existing limits from ATLAS~\cite{ATLAS:2024zxk, ATLAS:2022izj} (gray, solid/dotted) and CMS~\cite{CMS:2021sch} (gray, hatched) for $\mathcal{B}(h\to A'A') = 0.1\%$ (Top) and $0.01\%$ (Bottom). The ATLAS limits are set at $90\%$ CL.}
    \label{fig:ATLAS_vs_CMS_vs_FCC}
\end{figure}

The results are presented in Fig.~\ref{fig:ATLAS_vs_CMS_vs_FCC} where contours corresponding to $\mathcal{B}(h\to A'A') = 0.1\%$, $0.01\%$, and $0.005\%$ are shown. These are compared to existing limits from ATLAS~\cite{ATLAS:2024zxk,ATLAS:2022izj} and CMS~\cite{CMS:2021sch} at $0.1\%$ and $0.01\%$ respectively. In the bottom panel of Fig.~\ref{fig:ATLAS_vs_CMS_vs_FCC}, the ATLAS displaced search is absent because no interpretation below $\mathcal{B}(h\to A'A') = 0.1\%$ has been made public. We note that the ATLAS and CMS limits are set at different confidence levels, so the comparison should be taken as indicative.

\begin{figure}[ht!]
    \centering
    \includegraphics[width=0.48\linewidth]{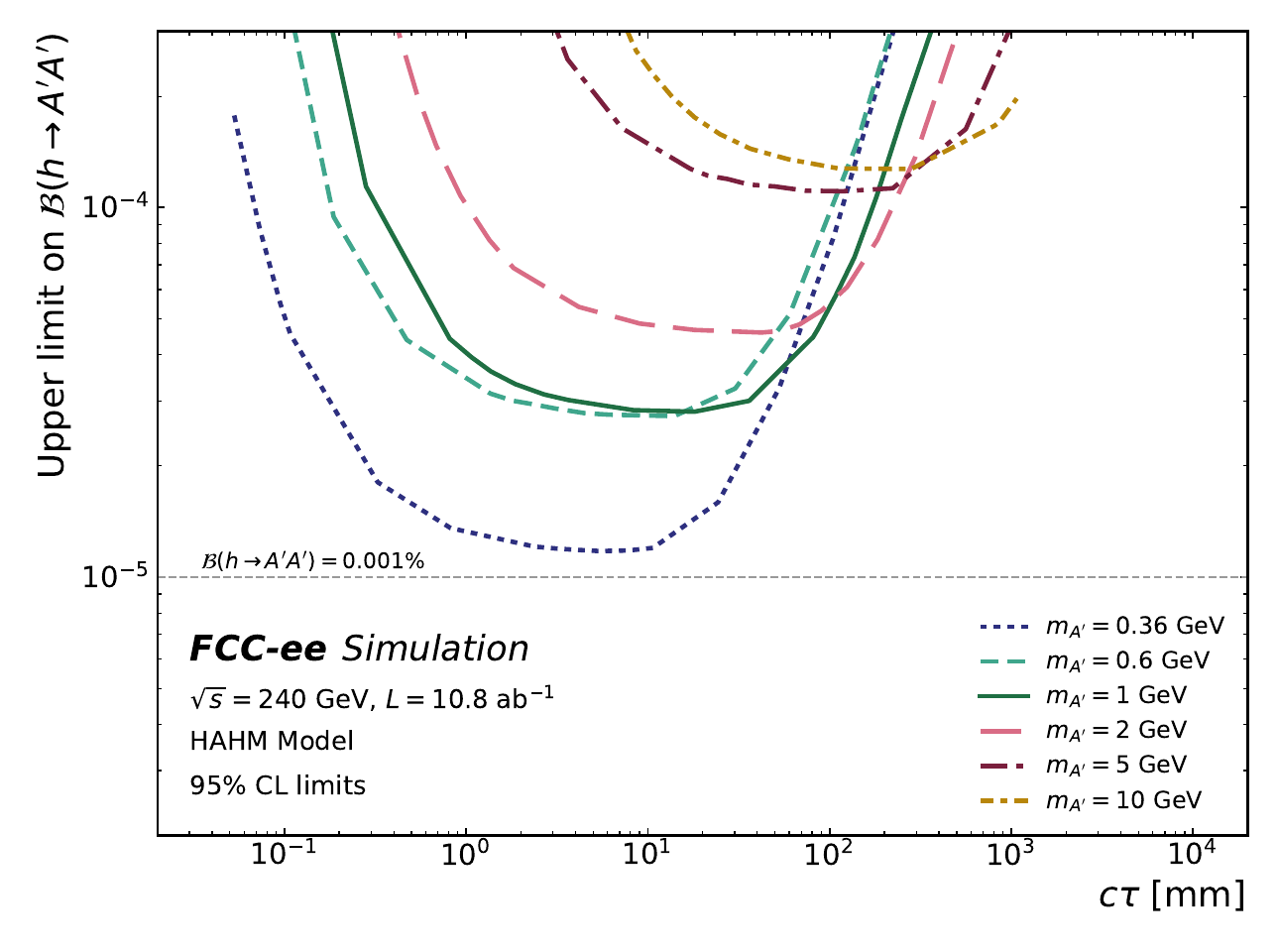}
        \includegraphics[width=0.48\linewidth]{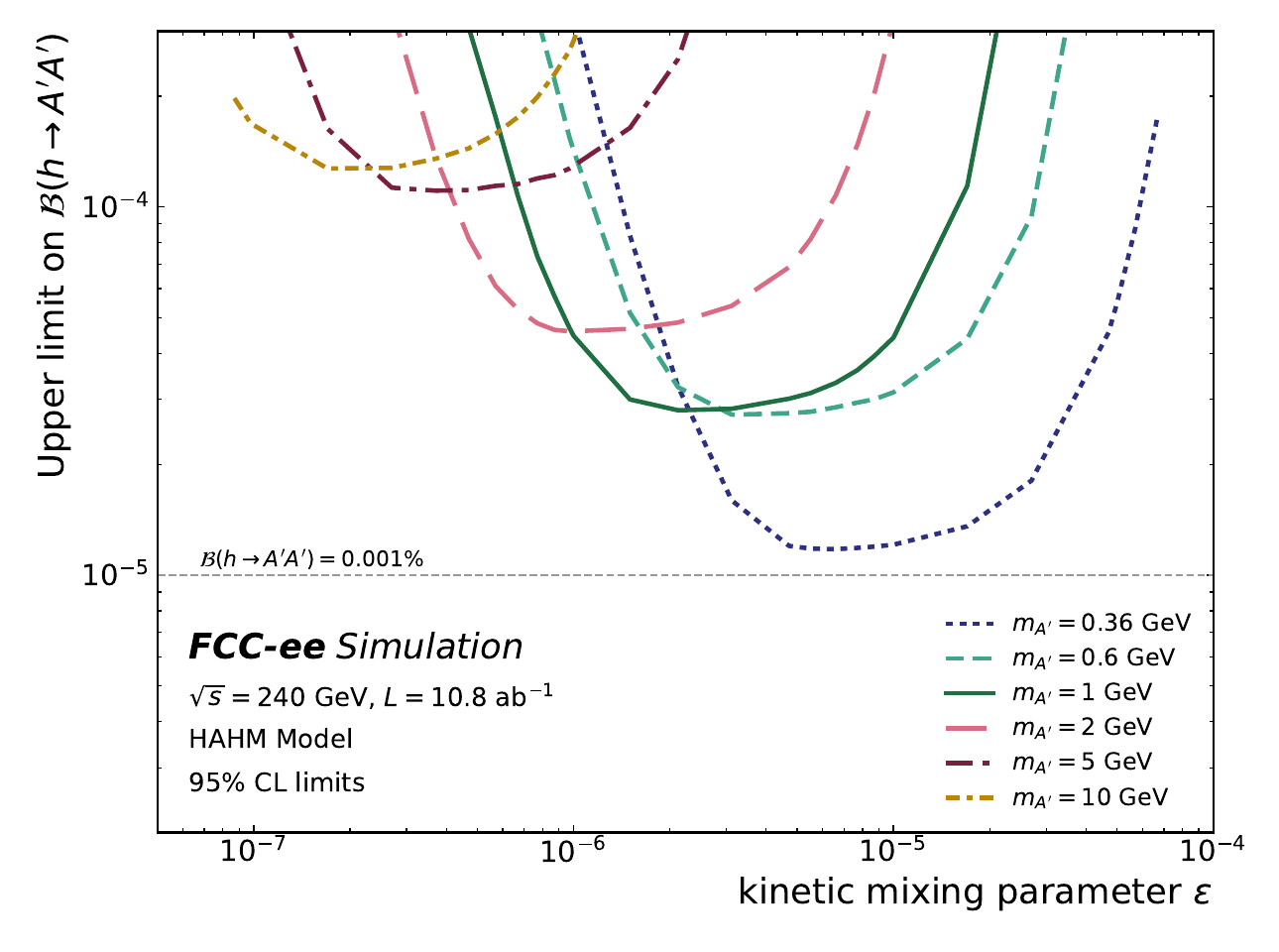}

    \caption{Exclusion limits at 95\% CL on the Higgs to dark photon branching fraction as a function of the dark photon mean proper lifetime, $c\tau$ (Left) and the kinetic mixing $\varepsilon$ (Right), for six different representative masses.}
    \label{fig:br_vs_ctau1}
\end{figure}

Fig.~\ref{fig:br_vs_ctau1} presents a summary of the $\mathcal{B}(h\to A'A')$ limits that could be observed at the FCC-ee as a function of both the dark photon mean proper lifetime, $c\tau$ and the kinetic mixing $\varepsilon$. The figure shows that the best reach for dark photons is at lifetimes of the order $\mathcal{O}(10-100)~$mm, for low-mass dark photons, and kinetic mixing on the order of $10^{-5}$.

\section{Conclusions}

In this work, we have presented a prospective study of exotic Higgs boson decays into dark photon pairs, $h \to A'A'$, at the electron-positron stage of the Future Circular Collider (FCC-ee) operating at $\sqrt{s}=240$~GeV. The analysis targets final states with two hadronic jets from $Z \to q\bar{q}$ and two pairs of collimated, displaced muons originating from long-lived dark photons.

Signal samples were generated within the framework of the Hidden Abelian Higgs Model, covering a broad parameter space in dark photon mass and kinetic mixing. Using a simple yet robust selection based on event kinematics and displaced vertex reconstruction, we demonstrate that the Standard Model background can be reduced to zero while maintaining a high signal efficiency over a wide region of parameter space.

In the absence of expected background, the sensitivity is driven by Poisson statistics, leading to an upper limit of approximately 3 signal events at 95\% confidence level. This allows us to set projected limits on the branching ratio $\mathcal{B}(h \to A'A')$ as a function of $m_{A'}$ and $\varepsilon$. The resulting sensitivity for a value of the Higgs to dark photon branching ratio of $0.1\%$ is comparable with published results from the ATLAS and CMS collaborations in three different searches. It is important to note that the search presented in this paper represents a first exploration of this kind of signature at FCC-ee with a very simple kinematic selection. The fact that this non-optimized search already covers current exclusions, obtained with more sophisticated strategies, indicates good potential, especially for lower branching ratios, more accessible at FCC-ee. 

In summary, this study demonstrates that FCC-ee provides sensitivity to dark sectors via exotic Higgs decays, with the potential to significantly constrain dark photons over a wide region of parameter space and low Higgs boson branching ratios.

\section{Acknowledgments}

This work was supported by the Swedish Research Council (Vetenskapsrådet) under grant agreement VR 2023-03403. The authors acknowledge the support of eSSENCE, a Swedish strategic research program in e-Science. 

The authors want to thank Giacomo Polesello for the discussions and guidance. 

\bibliography{bibliography.bib}

@article{Curtin:2013fra,
    author = "Curtin, David and others",
    title = "{Exotic decays of the 125 GeV Higgs boson}",
    eprint = "1312.4992",
    archivePrefix = "arXiv",
    primaryClass = "hep-ph",
    reportNumber = "YITP-13-47, PITT-PACC-1314",
    doi = "10.1103/PhysRevD.90.075004",
    journal = "Phys. Rev. D",
    volume = "90",
    number = "7",
    pages = "075004",
    year = "2014"
}

@article{FCC:2025lpp,
    author = "Benedikt, M. and others",
    collaboration = "FCC",
    title = "{Future Circular Collider Feasibility Study Report: Volume 1, Physics, Experiments, Detectors}",
    eprint = "2505.00272",
    archivePrefix = "arXiv",
    primaryClass = "hep-ex",
    reportNumber = "CERN-FCC-PHYS-2025-0002",
    doi = "10.1140/epjc/s10052-025-15077-x",
    journal = "Eur. Phys. J. C",
    volume = "85",
    number = "12",
    pages = "1468",
    year = "2025"
}

@article{Alimena:2019zri,
    author = "Alimena, Juliette and others",
    title = "{Searching for long-lived particles beyond the Standard Model at the Large Hadron Collider}",
    eprint = "1903.04497",
    archivePrefix = "arXiv",
    primaryClass = "hep-ex",
    doi = "10.1088/1361-6471/ab4574",
    journal = "J. Phys. G",
    volume = "47",
    number = "9",
    pages = "090501",
    year = "2020"
}

@article{deBlas:2025gyz,
    author = "de Blas, Jorge and others",
    title = "{Physics Briefing Book: Input for the 2026 update of the European Strategy for Particle Physics}",
    eprint = "2511.03883",
    archivePrefix = "arXiv",
    primaryClass = "hep-ex",
    reportNumber = "CERN--2025-008, CERN-ESU-2025-001",
    doi = "10.23731/CYRM-2025-008",
    month = "11",
    year = "2025"
}

@techreport{European:2957411,
      author        = "{The European Strategy Group}",
      title         = "{The European Strategy for Particle Physics: 2026 Update -
                       Deliberation document by the European Strategy Group}",
      reportNumber  = "CERN-ESU-2026-003",
      institution   = "CERN",
      address       = "Geneva",
      year          = "2026",
      url           = "https://cds.cern.ch/record/2957411",
      doi           = "10.17181/CERN.F5VS.K3VW",
}

@article{Alwall:2011uj,
    author = "Alwall, Johan and Herquet, Michel and Maltoni, Fabio and Mattelaer, Olivier and Stelzer, Tim",
    title = "{MadGraph 5 : Going Beyond}",
    eprint = "1106.0522",
    archivePrefix = "arXiv",
    primaryClass = "hep-ph",
    reportNumber = "FERMILAB-PUB-11-448-T",
    doi = "10.1007/JHEP06(2011)128",
    journal = "JHEP",
    volume = "06",
    pages = "128",
    year = "2011"
}

@article{Sjostrand:2014zea,
    author = {Sj{\"o}strand, Torbj{\"o}rn and Ask, Stefan and Christiansen, Jesper R. and Corke, Richard and Desai, Nishita and Ilten, Philip and Mrenna, Stephen and Prestel, Stefan and Rasmussen, Christine O. and Skands, Peter Z.},
    title = "{An introduction to PYTHIA 8.2}",
    eprint = "1410.3012",
    archivePrefix = "arXiv",
    primaryClass = "hep-ph",
    reportNumber = "LU-TP-14-36, MCNET-14-22, CERN-PH-TH-2014-190, FERMILAB-PUB-14-316-CD, DESY-14-178, SLAC-PUB-16122",
    doi = "10.1016/j.cpc.2015.01.024",
    journal = "Comput. Phys. Commun.",
    volume = "191",
    pages = "159--177",
    year = "2015"
}

@article{deFavereau:2013fsa,
    author = "de Favereau, J. and Delaere, C. and Demin, P. and Giammanco, A. and Lema{\^\i}tre, V. and Mertens, A. and Selvaggi, M.",
    collaboration = "DELPHES 3",
    title = "{DELPHES 3, A modular framework for fast simulation of a generic collider experiment}",
    eprint = "1307.6346",
    archivePrefix = "arXiv",
    primaryClass = "hep-ex",
    doi = "10.1007/JHEP02(2014)057",
    journal = "JHEP",
    volume = "02",
    pages = "057",
    year = "2014"
}

@article{Ilg:2025zri,
    author = "Ilg, Armin",
    title = "{The IDEA detector concept for FCC-ee}",
    eprint = "2510.26195",
    archivePrefix = "arXiv",
    primaryClass = "physics.ins-det",
    doi = "10.22323/1.485.0483",
    journal = "PoS",
    volume = "EPS-HEP2025",
    pages = "483",
    year = "2026"
}

@misc{WINTER2023,
    key = {},
    note = {},
    url={https://github.com/HEP-FCC/FCC-config/tree/winter2023}
}

@misc{FCCANALYSES,
  author       = {Helsens, Clement and
                  Perez, Emmanuel and
                  Selvaggi, Michele and
                  Volkl, Valentin and
                  Forthomme, Laurent and
                  {Munch Torndal}, Julie},
  title        = {{HEP-FCC/FCCAnalyses}: v0.12.0},
  year         = {2026},
  publisher    = {Zenodo},
  version      = {v0.12.0},
  doi          = {10.5281/zenodo.18229442},
  url          = {https://doi.org/10.5281/zenodo.18229442},
  note         = {Common analysis framework for the Future Circular Collider}
}

@article{Catani:1991hj,
    author = "Catani, S. and Dokshitzer, Yuri L. and Olsson, M. and Turnock, G. and Webber, B. R.",
    title = "{New clustering algorithm for multi - jet cross-sections in e+ e- annihilation}",
    reportNumber = "CAVENDISH-HEP-91-5",
    doi = "10.1016/0370-2693(91)90196-W",
    journal = "Phys. Lett. B",
    volume = "269",
    pages = "432--438",
    year = "1991"
}

@article{Curtin:2014cca,
    author = "Curtin, David and Essig, Rouven and Gori, Stefania and Shelton, Jessie",
    title = "{Illuminating Dark Photons with High-Energy Colliders}",
    eprint = "1412.0018",
    archivePrefix = "arXiv",
    primaryClass = "hep-ph",
    reportNumber = "YITP-SB-14-49",
    doi = "10.1007/JHEP02(2015)157",
    journal = "JHEP",
    volume = "02",
    pages = "157",
    year = "2015"
}

@article{Polesello:2025cbi,
    author = "Polesello, Giacomo",
    title = "{Sensitivity of the FCC-ee to the decay of a dark photon into a $\mu^+\mu^-$ pair}",
    eprint = "2511.23337",
    archivePrefix = "arXiv",
    primaryClass = "hep-ph",
    month = "11",
    year = "2025"
}

@article{Carceller:2025fjc,
    author = "Carceller, Juan Miguel and Fila, Mateusz Jakub and Francois, Brieuc and Gaede, Frank and Hegner, Benedikt and Madlener, Thomas and Smiesko, Juraj and Sailer, Andr{\'e}",
    title = "{EDM4hep - The common event data model for the Key4hep project}",
    doi = "10.1051/epjconf/202533701131",
    journal = "EPJ Web Conf.",
    volume = "337",
    pages = "01131",
    year = "2025"
}

@article{Dam:2025zed,
    author = "Dam, Mogens",
    title = "{Detector requirements, design, and technologies for the FCC-ee Higgs, electroweak, and top factory}",
    eprint = "2505.06781",
    archivePrefix = "arXiv",
    primaryClass = "hep-ex",
    doi = "10.1016/j.nima.2025.170648",
    journal = "Nucl. Instrum. Meth. A",
    volume = "1080",
    pages = "170648",
    year = "2025"
}

@article{LCFIplus,
   title={LCFIPlus: A framework for jet analysis in linear collider studies},
   volume={808},
   ISSN={0168-9002},
   url={http://dx.doi.org/10.1016/j.nima.2015.11.054},
   DOI={10.1016/j.nima.2015.11.054},
   journal={Nuclear Instruments and Methods in Physics Research Section A: Accelerators, Spectrometers, Detectors and Associated Equipment},
   publisher={Elsevier BV},
   author={Suehara, Taikan and Tanabe, Tomohiko},
   year={2016},
   month=Feb, pages={109–116} }

@article{ATLAS:2024zxk,
    author = "Aad, Georges and others",
    collaboration = "ATLAS",
    title = "{Search for light neutral particles decaying promptly into collimated pairs of electrons or muons in pp collisions at $\sqrt{s}$ = 13 $\text {T}\text {e}\hspace{-1.00006pt}\text {V}$ with the ATLAS detector}",
    eprint = "2407.09168",
    archivePrefix = "arXiv",
    primaryClass = "hep-ex",
    reportNumber = "CERN-EP-2024-183",
    doi = "10.1140/epjc/s10052-025-13916-5",
    journal = "Eur. Phys. J. C",
    volume = "85",
    number = "3",
    pages = "335",
    year = "2025"
}

@article{ATLAS:2022izj,
    author = "Aad, Georges and others",
    collaboration = "ATLAS",
    title = "{Search for light long-lived neutral particles that decay to collimated pairs of leptons or light hadrons in pp collisions at $ \sqrt{s} $ = 13 TeV with the ATLAS detector}",
    eprint = "2206.12181",
    archivePrefix = "arXiv",
    primaryClass = "hep-ex",
    reportNumber = "CERN-EP-2022-054",
    doi = "10.1007/JHEP06(2023)153",
    journal = "JHEP",
    volume = "06",
    pages = "153",
    year = "2023"
}

@article{CMS:2021sch,
    author = "Tumasyan, Armen and others",
    collaboration = "CMS",
    title = "{Search for long-lived particles decaying into muon pairs in proton-proton collisions at $ \sqrt{s} $ = 13 TeV collected with a dedicated high-rate data stream}",
    eprint = "2112.13769",
    archivePrefix = "arXiv",
    primaryClass = "hep-ex",
    reportNumber = "CMS-EXO-20-014, CERN-EP-2021-266",
    doi = "10.1007/JHEP04(2022)062",
    journal = "JHEP",
    volume = "04",
    pages = "062",
    year = "2022"
}

@article{ParticleDataGroup:2024cfk,
    author = "Navas, S. and others",
    collaboration = "Particle Data Group",
    title = "{Review of particle physics}",
    doi = "10.1103/PhysRevD.110.030001",
    journal = "Phys. Rev. D",
    volume = "110",
    number = "3",
    pages = "030001",
    year = "2024"
}

\end{document}